\documentclass[twocolumn]{aastex62}
\usepackage{amssymb}
\usepackage{array,multirow}
\usepackage{comment}
\usepackage{enumerate}
\usepackage{bm}

\newcommand{\gaia}{{\it Gaia}}

\newcommand{\Porb}{\ifmmode {P_{\rm orb}}\else${P_{\rm orb}}$\fi}

\newcommand{\Msun}{\ifmmode {{M_\odot}}\else{$M_\odot$}\fi}
\newcommand{\Mtot}{\ifmmode {{M_{\rm tot}}}\else{$M_{\rm tot}$}\fi}
\newcommand{\RV}{\ifmmode {{\rm RV}}\else RV \fi}
\newcommand{\bigG}{\ifmmode {\mathcal{G}}\else${\mathcal{G}}$\fi}

\submitjournal{ApJ}

\shorttitle{Weighing the Darkness}

\begin{document}

\title{Weighing the Darkness III: How \gaia\ Could, but Probably Won't, Astrometrically Detect Free-Floating Black Holes}

\author[0000-0001-5261-3923]{Jeff J. Andrews}
\affiliation{Center for Interdisciplinary Exploration and Research in Astrophysics (CIERA), 
1800 Sherman Ave., 
Evanston, IL, 60201, USA}
\affiliation{Department of Physics, University of Florida, 2001 Museum Rd., Gainesville, FL 32611}
\email{jeffrey.andrews@northwestern.edu}

\begin{abstract}
The gravitational pull of an unseen companion to a luminous star is well-known to cause deviations to the parallax and proper motion of a star. In a previous paper in this series, we argue that the astrometric mission \gaia\ can identify long-period binaries by precisely measuring these arcs. An arc in a star's path can also be caused by a fly-by -- the hyperbolic encounter with another massive object. We quantify the apparent acceleration over time induced by a companion star as a function of the impact parameter, velocity of interaction, and companion mass. In principle, \gaia\ could be used to astrometrically identify the contribution of massive compact halo objects to the local dark matter potential of the Milky Way. However, after quantifying their rate and \gaia's sensitivity, we find that fly-bys are so rare that {\it Gaia} will probably never observe one. Therefore every star in the {\it Gaia} database exhibiting astrometric acceleration is likely in a long-period binary with another object. Nevertheless, we show how intermediate mass black holes, if they exist in the Solar Neighborhood, could be detected by the anomalously large accelerations they induce on nearby stars.
\end{abstract}


\section{Introduction}
\label{S:intro}

There hardly exists an area of stellar astrophysics unaffected by the \gaia\ astrometric mission. The latest third data release contains over 1.8 billion stars, of which 1.5 billion have measured parallaxes and proper motions, a factor of $10^4$ increase over the previous HIPPARCOS catalog \citep{Gaia_mission, lindegren2021, gaia_EDR3}. The combined improvement in astrometric precision and catalog size has allowed for new, previously unthinkable measurements \citep[for a review, see][]{brown2021} including the detection of waves in the local Milky Way stellar population \citep{bennett2019}, the discovery of low-density, coherent stellar structures \citep{kounkel2019}, and the precise dynamics of Milky Way globular clusters \citep{vasiliev2019}. 

With its third data release containing $\sim10^5$ astrometrically resolved binary orbits, \gaia's impact on binary astrophysics is particularly influential. Such astrometric orbits provide some of the most precise measurements of the masses of stellar and substellar companions to stars \citep{Hartkopf1996, Mason1999, Perryman2014, Sozzetti2014, andrews2019, brandt2019, penoyre2020}. With the release of the HIPPARCOS data set it was realized that, if sufficiently widely separated, a massive companion can induce non-linear proper motions---astrometric accelerations---on a star \citep{Wielen1997, Pourbaix2000}. 

By comparing data from HIPPARCOS with the second data release from \gaia, \citet{Brandt2018} \citep[which has since been updated to the \gaia\ early data release 3;][]{brandt2021a} and \citet{kervella2019a} have both separately released catalogs of stars exhibiting astrometric accelerations. These catalogs have been used for follow-up to efficiently find new brown dwarfs \citep{bowler2021, bonavita2022, kuzuhara2022} and exoplanets \citep{errico2022} and to dynamically characterize previously known systems \citep{brandt2021b, feng2021, li2021, steiger2021, zeng2021, biller2022, dupuy2022, franson2022}.

In addition to stellar and exoplanet companions, some fraction of these astrometric binaries may host compact object companions, both in resolved orbits \citep{gould2002, Tomsick2010, Barstow2014, Breivik2017, Mashian2017, Yalinewich2018, Yamaguchi2018, Breivik2018, andrews2019, chawla2021} and in long-period binaries where only partial arcs are observed \citep{andrews2021}. However, astrometric accelerations from long-period orbits could also be induced by fly-by encounters with dark, massive objects.

Previous methods to detect the existence and rate of free-floating massive objects in the Milky Way are limited to microlensing detections \citep{paczynski86, alcock1993}. With photometry alone, degeneracies in the signals preclude definitive detections of individual systems although population conclusions can still be made \citep{wyrzykowski2020, golovich2022}. If a perturbation in the position of a star is simultaneously measured, so-called astrometric microlensing events can allow for a measurement of the mass, distance, and proper motion of free-floating black holes \citep{lam2022, sahu2022}. \citet{andrews2022} used one such detection to constrain the kicks that BHs receive at birth \citep[see also][]{vigna-gomez2022}. While microlensing experiments have identified samples of black hole candidates \citep[e.g.,][]{bennett2002, mao2002}, studies of wide binaries and microlensing populations both indicate that the overall composition of dark objects is unlikely to comprise a large fraction of the overall gravitational potential of the Milky Way \citep{bahcall85, weinberg87, tisserand07, yoo04, tian19}. 

This work is the third in a series exploring the possibilities of using the precise astrometry provided by \gaia\ for identifying dark companions to luminous stars. In the first paper of this series \citep{andrews2019} we quantify \gaia's ability to measure the masses of the dark companions in such orbits. In the second paper of this series \citep[hereafter Paper II;][]{andrews2021}, we extend our analysis to quantify \gaia's ability to detect astrometric acceleration from stars in very wide binaries with periods longer than \gaia's lifetime. In this work, we apply our results quantifying \gaia's detection sensitivity for astrometric acceleration to the detection of fly-by encounters from dark, massive objects.

In Section~\ref{sec:hyperbolic_orbits} we discuss the characteristics of hyperbolic orbits describing fly-by interactions, and in Section~\ref{sec:rates} we calculate the number of fly-by interactions that \gaia\ can detect under the extreme assumption that all dark matter is comprised of massive compact halo objects (MACHOs). In Section~\ref{sec:IMBH} we quantify \gaia's ability to detect intermediate mass black holes, if one were to exist in the Solar Neighborhood. We provide some discussion and conclusions in Section~\ref{sec:conclusions}.

\section{Hyperbolic Orbits Revisited}
\label{sec:hyperbolic_orbits}

Since for the orbits in question we observe only a fraction of the entire interaction, an orbit's eccentricity, and therefore its status as a bound or unbound orbit, is not immediately apparent. The interaction between such unbound stars are ubiquitous in the Milky Way; indeed the Sun is being pulled - however slightly - by nearby stars in the Solar Neighborhood. To actually determine the typical fly-by interactions that \gaia\ will observe, we first review the relevant equations describing hyperbolic orbits. 

\begin{figure}
    \begin{center}
    \includegraphics[width=1.0\columnwidth]{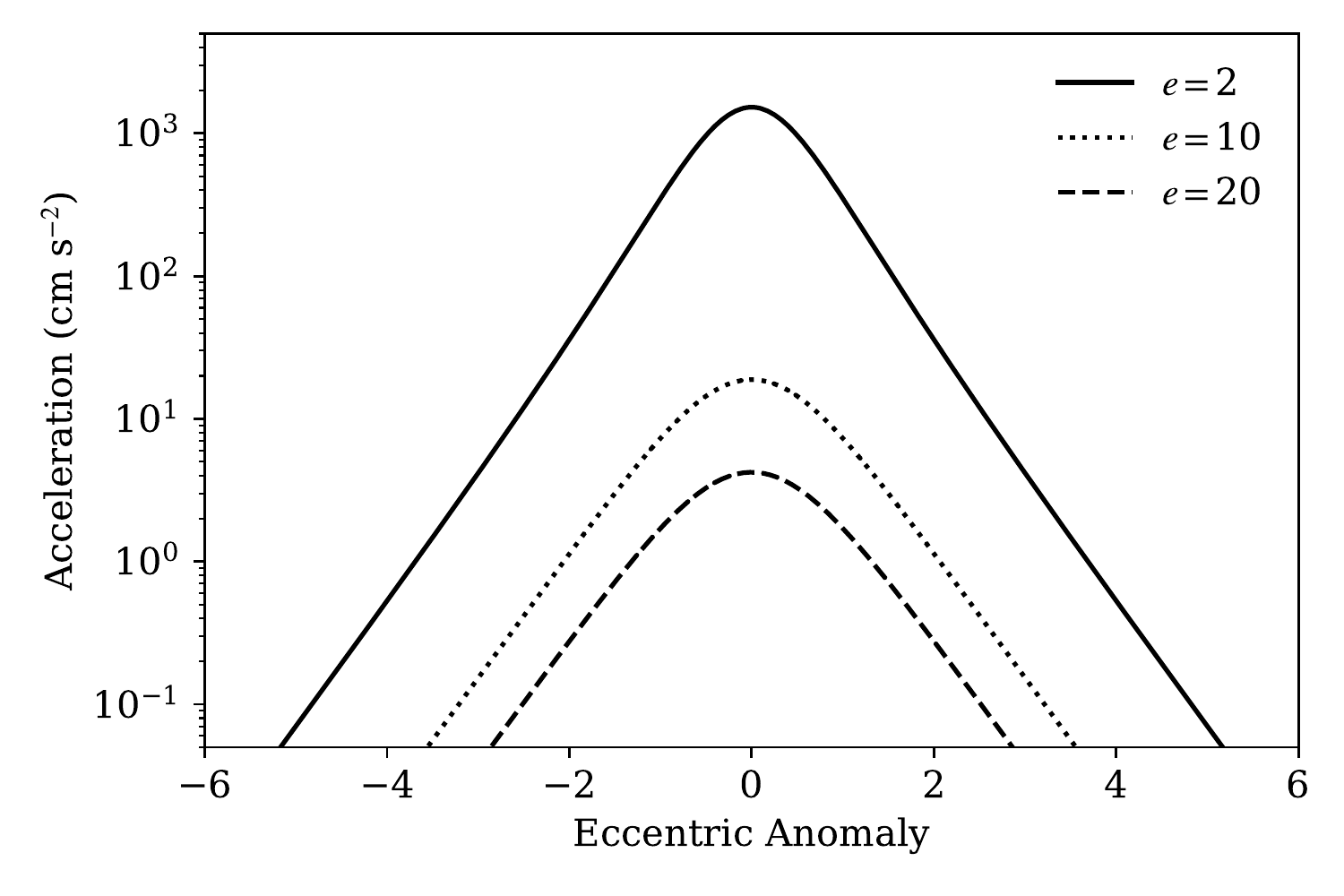}
    \caption{ The acceleration felt by a star as a function of eccentric anomaly for fly-bys with different eccentricities. Acceleration is maximized at the closest approach, near pericenter, and for orbits with the lowest $e$ (and therefore the smallest impact parameter). Note that, contrary to the case of elliptical orbits, for fly-by encounters, $E$ has a domain ranging from $-\infty$ to $\infty$. }  
    \label{fig:hyperbolic_accelerations}
    \end{center}
\end{figure}

For a hyperbolic encounter with a semi-major axis, $a$, and an eccentricity $e$, the separation of two stars, $r$, can be expressed as:
\begin{equation}
    r = a (e \cosh E - 1),
    \label{eq:r}
\end{equation}
where $E$ is the eccentric anomaly. Contrary to elliptical orbits where $0<E<2\pi$, in hyperbolic orbits $-\infty < E < \infty$. For a fly-by interaction, the semi-major axis is set by the two objects' masses, $M_1$ and $M_2$ and the fly-by velocity, $V_{\infty}$:
\begin{equation}
    a=\mathcal{G}(M_1+M_2) / V_{\infty}^2.
    \label{eq:a}
\end{equation}
Similarly, the eccentricity $e$ can be determined from the ratio of the impact parameter, $b$, to the semi-major axis:
\begin{equation}
    e = \sqrt{b^2/a^2 + 1}.
\end{equation}
Note that for sufficiently large impact parameters, $e\approx b/a$.

The acceleration from a passing star with mass $M_2$ felt by an observed star of mass $M_1$ is:
\begin{equation}
    \frac{d^2x_1}{dt^2} = -\frac{\mathcal{G}M_2}{a^2} \frac{1}{(e \cosh E - 1)^2}.
    \label{eq:accel}
\end{equation}
In the limit of large eccentricity, this can be reduced to:
\begin{equation}
    \lim_{e \gg 1} \frac{d^2x_1}{dt^2} = -\frac{\mathcal{G}M_2}{b^2} \frac{1}{\cosh^2 E}
\end{equation}

To obtain a distribution of accelerations observed for a fly-by encounter, we recall that for a particular hyperbolic eccentricity, we can express the mean anomaly, $M$, as:
\begin{equation}
    M = \sqrt{\frac{\mathcal{G} (M_1+M_2)}{a^3}} (t - \tau/2),
    \label{eq:M}
\end{equation}
where $t$ is the time along an orbit (where $t=0$ corresponds to the pericenter) and $\tau$ is the maximum time we are considering for a particular interaction. This corresponds to a minimum $M$, $M_{\rm min}$, which occurs at $-\tau/2$ and a maximum $M$, $M_{\rm max}$, which occurs at $\tau/2$. Since $P(M) \sim 1$:
\begin{equation}
    P(M) = \frac{1}{M_{\rm max} - M_{\rm min}} = \sqrt{\frac{a^3}{\mathcal{G} (M_1+M_2)}} \frac{1}{\tau},
    \label{eq:prob_M_hyperbolic}
\end{equation}
for $M_{\rm min} < M < M_{\rm max}$. Since $P(E) = P(M) | \partial M / \partial E |$ and $M = e \sinh E - E$, we can determine the probability of finding a fly-by encounter with eccentricity $e$ at a particular $E$:
\begin{equation}
    P(E | e) = \sqrt{\frac{a^3}{\mathcal{G} (M_1+M_2)}} \frac{1}{\tau} (e \cosh E - 1).
    \label{eq:prob_E_hyperbolic}
\end{equation}

To find the distribution of accelerations exhibited by a single hyperbolic encounter with eccentricity $e$, we multiply $P(E | e)$ by a Jacobian term:
\begin{equation}
    P(d^2x_1/dt^2 | e) = P(E^*|e)\ \left| \frac{\partial E}{\partial (d^2x_1/dt^2)} \right|_{E=E^*},
    \label{eq:prob_hyperbolic_acceleration}
\end{equation}
where $E^*$ is the eccentric anomaly corresponding to an acceleration, $d^2x_1/dt^2$. Solving Equation~\ref{eq:accel}, we find:
\begin{equation}
    E^* = \cosh^{-1} \left\{ \frac{1}{e} + \frac{1}{e} \left[-\frac{\mathcal{G}M_2}{a^2} \left(\frac{d^2x_1}{dt^2}\right)^{-1}\right]^{1/2} \right\}.
    \label{eq:calc_Estar}
\end{equation}
The Jacobian term in Equation~\ref{eq:prob_hyperbolic_acceleration} can be calculated by taking the derivative of $d^2x_1/dt^2$ with respect to $E$:
\begin{equation}
    \frac{\partial (d^2x_1/dt^2)}{\partial E} = \frac{2 \mathcal{G}M_2}{a^2} \frac{e \sinh E}{(e \cosh E - 1)^3}.
\end{equation}

\begin{figure}
    \begin{center}
    \includegraphics[width=1.0\columnwidth]{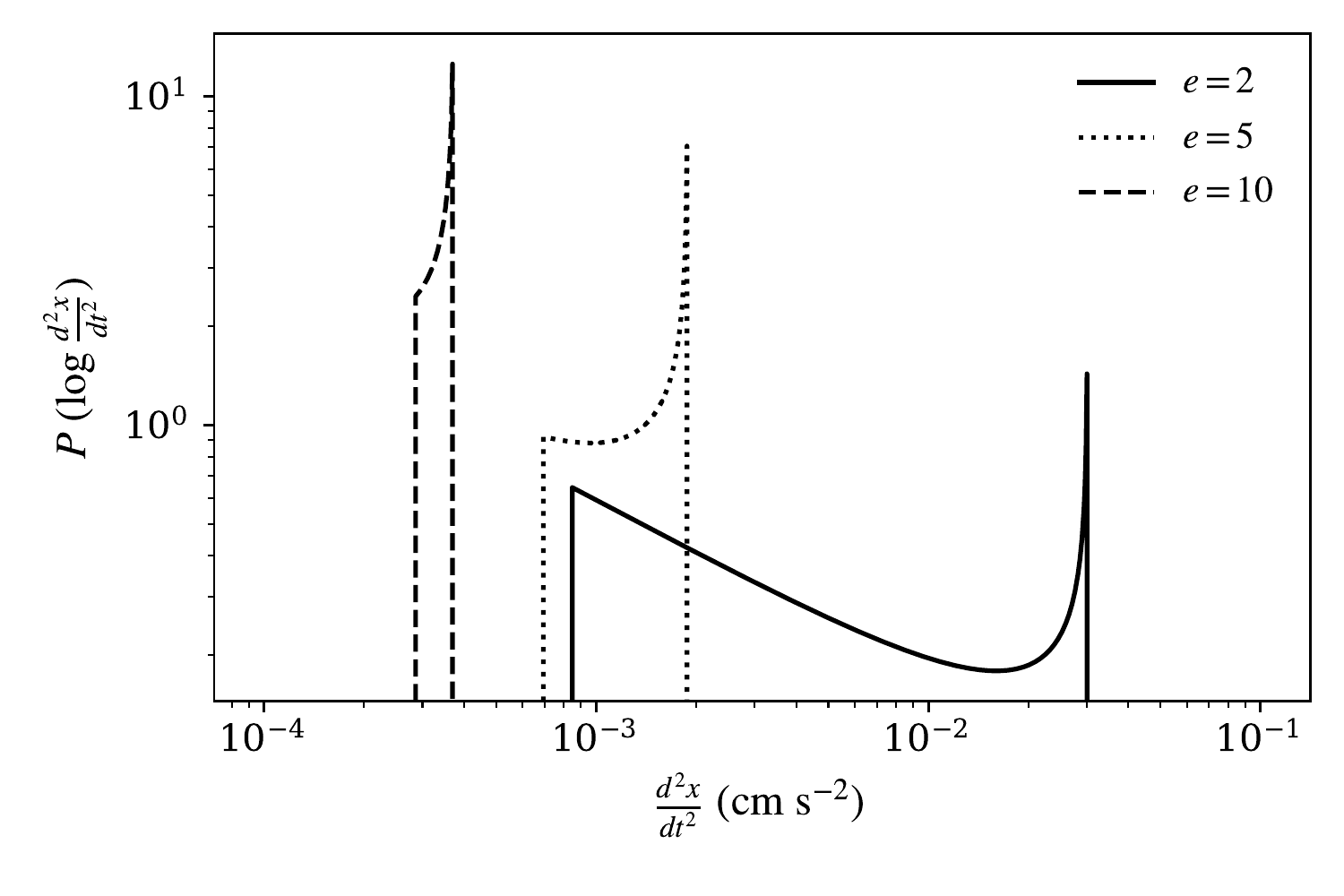}
    \caption{ The orbital acceleration felt by a star with the length of time it spends at each part of a hyperbolic orbit over a 10-year interaction time around orbital pericenter. These accelerations are calculated for two 1 \Msun\ stars, interacting at $V_{\infty}=20$ km s$^{-1}$. Since the overall interaction time takes longer for higher eccentricity fly-bys, the $e=10$ orbit shows less dynamic range over a fixed lifetime.}  
    \label{fig:prob_hyperbolic_acceleration}
    \end{center}
\end{figure}

Figure~\ref{fig:prob_hyperbolic_acceleration} shows the distribution of accelerations felt by a 1 \Msun\ star with a 1 \Msun\ fly-by with a velocity of $V_{\infty}=20$ km s$^{-1}$ for three different choices of orbital eccentricity (or equivalently different impact parameters) over a 10-year interaction time, calculated from Equation~\ref{eq:prob_hyperbolic_acceleration}. There is a clear spike in the distribution at relatively high accelerations around pericenter, which can be obtained by setting $E=0$ in Equation \ref{eq:accel}. Figure \ref{fig:prob_hyperbolic_acceleration} also shows a progressively increasing probability at lower accelerations. The low acceleration cut-off in this figure is somewhat artificial since we only focus on the orbital phase within five years either side of pericenter. Since the overall interaction takes longer for orbits with higher eccentricities, these orbits show less dynamic range over a fixed lifetime. 

To obtain a complete distribution of accelerations, we must convolve over all impact parameters, $b$, or at least to some sufficiently large impact parameter, $b_{\rm max}$, such that any more distant fly-bys are well beyond observational detectability:
\begin{equation}
    P(d^2x_1/dt^2) = \int_0^{b_{\rm max}}\ db\ P(b)\ P(d^2x_1/dt^2 | b).
    \label{eq:prob_accel}
\end{equation}
The first term in the integrand is simply $P(b) = 2b/b_{\rm max}^2$, as the probability of an encounter with impact parameter $b$ scales with the differential probability of the area of the corresponding annulus. The second term in the integrand can be calculated from the expression in Equation~\ref{eq:prob_hyperbolic_acceleration}, where $e$ can be straightforwardly calculated from $a$ and $b$.

Finally, the probability of any fly-by encounter having an acceleration greater than some detectable limit, can be obtained by integrating Equation~\ref{eq:prob_accel} from the detection limit to some large number. In practice, the upper integration limit has little effect on the result, as most of the probability volume lies very close to the detection limit.

\section{The Rate of Fly-Bys}
\label{sec:rates}
\begin{figure}
    \begin{center}
    \includegraphics[width=1.0\columnwidth]{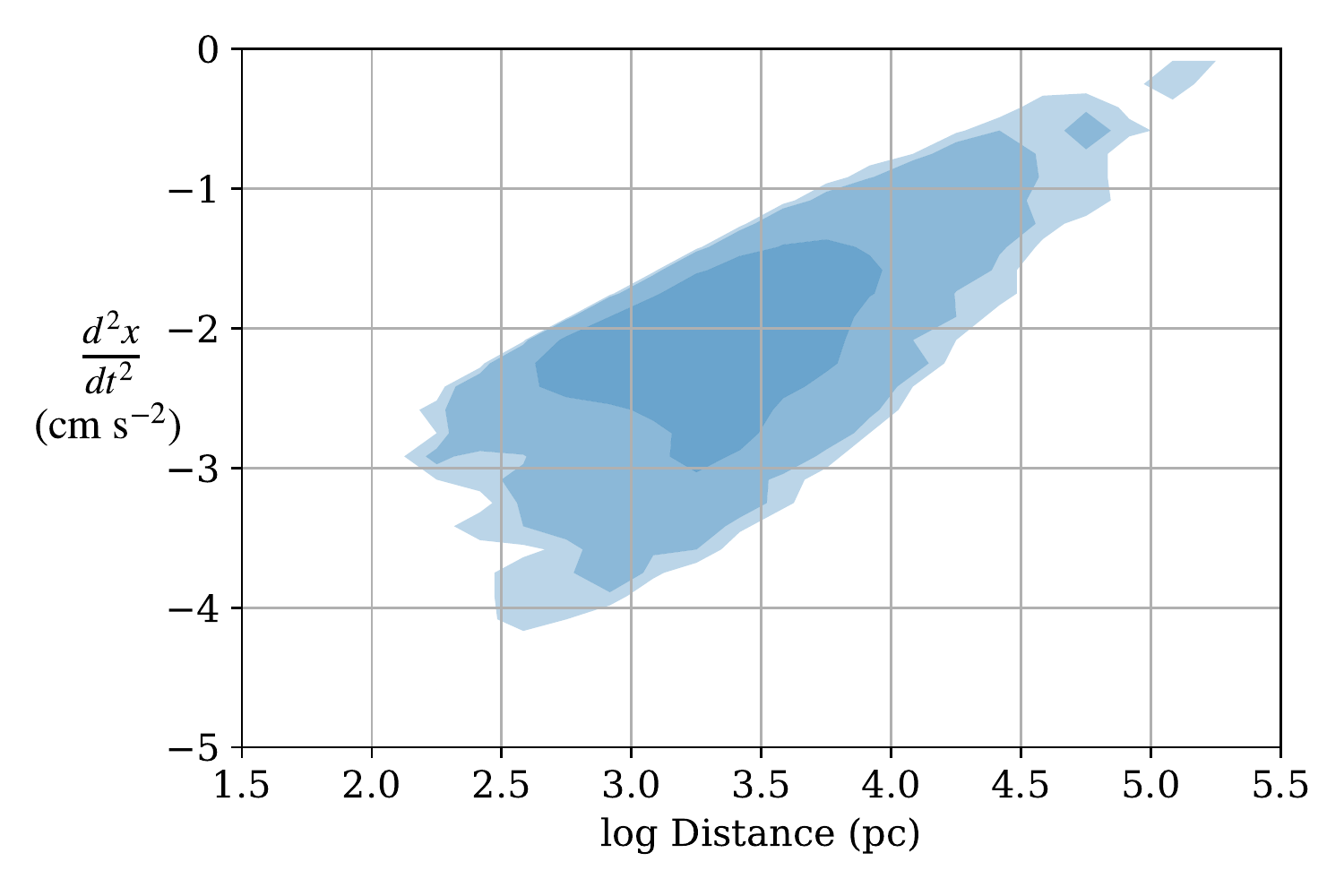}
    \caption{ The limiting accelerations for stars in the Gaia catalog as a function of distance. Stars at a 1 kpc distance have accelerations detectable to better than $\simeq 10^{-2}$~cm~s$^{-2}$, while even closer stars have detectable accelerations better than $10^{-4}$ cm s$^{-2}$. }
    \label{fig:Gaia_sample}
    \end{center}
\end{figure}

For an individual star, the average time between fly-bys can be approximated as $\Delta t = 1 / n \sigma V_{\infty}$, where $n$ is the number density of external perturbers, $\sigma=\pi b^2_{\rm max}$ is the cross-section of interaction, and $V_{\infty}$ is the relative velocity of interaction. The number of interactions over some time $\tau$ can be calculated as $n \sigma V_{\infty} \tau$. Note that the interaction time here is the same interaction time in Equation~\ref{eq:prob_E_hyperbolic}. For a single, $i$th star, the number of detectable fly-by interactions it will feel is the number of interactions it feels multiplied by the likelihood that any one interaction is above some critical acceleration indicating detectability:
\begin{equation}
    N_i = n \sigma V_{\infty} \tau\ P(d^2x_1/dt^2 > d^2x_1/dt^2_{{\rm crit},i}).
    \label{eq:N}
\end{equation}

How then to choose $\tau$ and $b_{\rm max}$? Combining all the terms to calculate $N$, one finds that the $\tau$ in Equation~\ref{eq:N} cancels out with the $\tau$ in Equation~\ref{eq:prob_E_hyperbolic}. Likewise, $b_{\rm max}^2$ term in $\sigma$ cancels with the $b_{\rm max}^2$ term in $P(b)$. Therefore, the calculation for $N$ is independent of these both terms. In practice, one needs to numerically calculate the integral in Equation~\ref{eq:prob_accel}, and therefore a judicious choice must be made for both $\tau$ and $b_{\rm max}$, such that they are sufficiently large to encompass all detectable fly-by interactions. 

\begin{figure}
    \begin{center}
    \includegraphics[width=1.0\columnwidth]{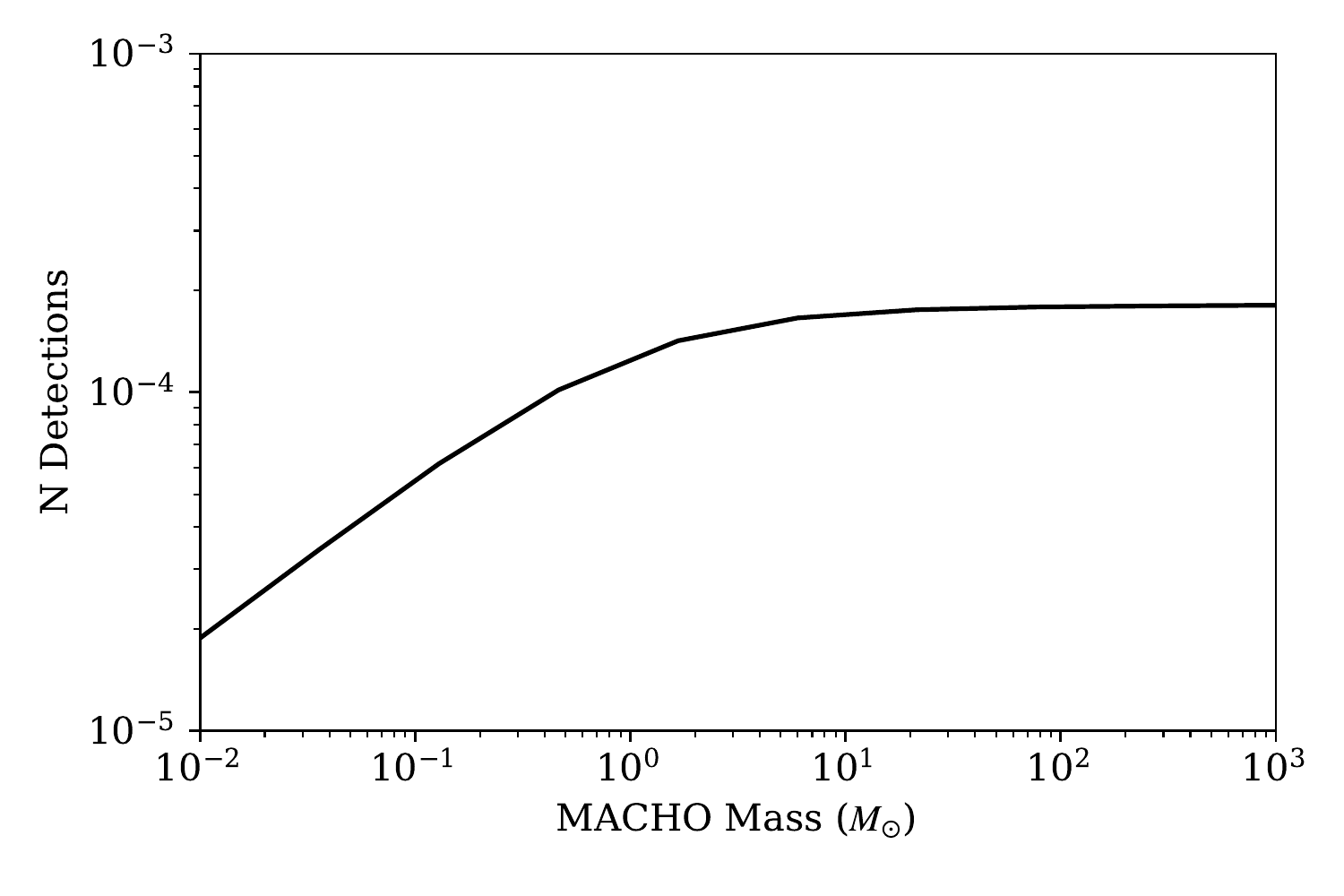}
    \caption{ Under the limiting assumption that all dark matter is comprised of MACHOs, we show the number of stars that will exhibit an acceleration detectable by Gaia, as a function of MACHO mass. There is a small fraction of a chance that one of the $\sim10^5$ accelerating signals identified by Gaia is caused by a MACHO; however, we would be likely unaware if one is detected as the accelerations produced by a MACHO are identical to those from exoplanets or faint stellar companions in bound binaries. }
    \label{fig:N_detections}
    \end{center}
\end{figure}

In paper II, we fit mock \gaia\ observations to derive the measurement precision for acceleration:
\begin{equation}
    \xi = \frac{\sigma (d^2x_1/dt^2)}{d^2x_1/dt^2} \simeq 35 \frac{d R^2}{\mathcal{G}M_2}\frac{\sigma_{\rm G}}{\tau_{\rm G}^2} \sqrt{\frac{1}{N_{\rm G}}}
    \label{eq:detectability_empirical},
\end{equation}
where $d$ is the distance to the star, $\sigma_{\rm G}$ is the astrometric precision for an individual measurement of a star, $\tau_{\rm G}$ is the lifetime of \gaia, and $N_{\rm G}$ is the number of observations of an individual star\footnote{In this work we are only considering the possibility of detecting stellar accelerations with astrometry, not radial velocity which has been explored elsewhere \citep{ravi2019, silverwood_easther2019, chakrabarti2020}.}. We follow the same procedure as in Paper II for calculating $\sigma_{\rm G}$. We further assume a \gaia\ lifetime of 10 years and $N_{\rm G}$ of 140 separate observations per star. Setting $\xi$ to 0.1, we can solve for the limiting acceleration detectable by an individual star. Figure~\ref{fig:Gaia_sample} shows the distribution of limiting accelerations as a function of distance. For the bulk of the distribution at distances of $\simeq$1 kpc, \gaia\ can detect accelerations down to $10^{-2}$ cm s$^{-2}$, while that improves to $10^{-4}$ cm s$^{-2}$ or better for stars in the nearest 100 pc.

As an extremely optimistic assumption for the possibility of fly-by interactions, we consider that all dark matter is comprised of MACHOs. We therefore set $n$ in Equation~\ref{eq:N} to $\rho_{\rm DM} / M_{\rm MACHO}$, where $\rho_{\rm DM}=0.01 \Msun\ {\rm pc}^{-3}$ is the local density of dark matter \citep{read2014} and $M_{\rm MACHO}$ is the mass of individual MACHO objects. We can further adopt $V_{\infty}=300$ km s$^{-1}$ to approximate the interaction speed with thin disk stars. Even under this optimistic scenario, any individual star is unlikely to be sufficiently close to a putative MACHO to exhibit a detectable acceleration. However, the DR3 catalog in \gaia\ contains 1.8 billion stars. To determine the overall number of detectable accelerations due to a MACHO model for dark matter, we separately calculate $N_i$ for all stars in the \gaia\ catalog and sum over them:
\begin{equation}
    N_{\rm Tot} = \sum_{\gaia} N_i.
\end{equation}
For computational efficiency, we do not sum over every star in the \gaia\ catalog, but extrapolate from a subset of 10,000 randomly selected stars.

Figure~\ref{fig:N_detections} shows the number of detectable accelerating signals for a full 10-year \gaia\ mission as a function of the dark object mass, under the optimistic assumption that all of dark matter is comprised of MACHOs. The distribution saturates at large MACHO masses having a probability of $10^{-4}$ of having a single detection, while the probability decreases at lower masses. It is clear that even a single detection is extremely unlikely. Since dark matter sets the upper limit on the local density of possible perturbers, it is therefore unlikely that any of the observed accelerating signals will be due to a fly-by interaction. We conclude that virtually all the accelerations detected by \gaia\ are due to bound objects: compact objects or faint stellar or substellar companions.

\section{Intermediate Mass Black Holes}
\label{sec:IMBH}

Even though they are unlikely to make up a significant portion of the overall Milky Way potential, IMBHs still may exist in the Solar Neighborhood. With the detection of a gravitational wave merger by LIGO and Virgo which formed a 142~$\Msun$ BH \citep{GW190521} as well as electromagnetic observations of high-luminosity objects such as HLX-1 \citep{davis2011, webb2012}, evidence for the existence of intermediate mass black holes (IMBHs) is strengthening. Previous methods to detect IMBHs that do not rely upon model-dependent electromagnetic observations include searching gravitational microlensing events \citep{mirhosseini2018, kains2018, blaineau2022, franco2021} and lensed gravitational wave events \citep{lai2018, gais2022} and gamma-ray bursts \citep{paynter2021}. Astrometric acceleration offers a previously unexplored method to indirectly detect the existence of an IMBH in the Solar Neighborhood. 

The acceleration caused by an IMBH of mass $M_{\rm IMBH}$ at separation $R$ from a luminous star is $\sim \mathcal{G} M_{\rm IMBH} / R^2$:
\begin{equation}
    \frac{d^2x_1}{dt^2} \sim 1.5 \times 10^{-5} \left( \frac{M_{\rm BH}}{10^5~M_{\odot}} \right) \left( \frac{R}{1~{\rm pc}} \right)^{-2} \ {\rm cm}\ {\rm s}^{-2}.
    \label{eq:accel_IMBH}
\end{equation}

Using this scaling, a $10^3~\Msun$ BH will need to wander within 2000~a.u.\ of a nearby, well-measured luminous star to be detectable. Such systems are quite rare; if they exist, they are only detectable within the nearest $\sim$10 pcs. Furthermore, such acceleration is hardly an unambiguous indicator of the presence of an IMBH. The degeneracy between mass and separation means that planets, faint stars, or even stellar mass BHs can all produce acceleration signals of similar magnitude. For instance, the same acceleration could also be caused by a 0.05~$\Msun$ brown dwarf at a separation of 17~a.u.\ in a bound orbit\footnote{While an Earth-mass exoplanet at a separation of 0.1~a.u.\ can produce the same acceleration, it is unlikely to be confused with a $10^3$~$\Msun$ IMBH, as the exoplanet would complete an orbit over Gaia's observational lifetime.}. Since brown dwarfs are far more numerous than IMBHs they are the far more likely culprit for any individual acceleration signature.

\begin{figure}
    \begin{center}
    \includegraphics[width=1.0\columnwidth]{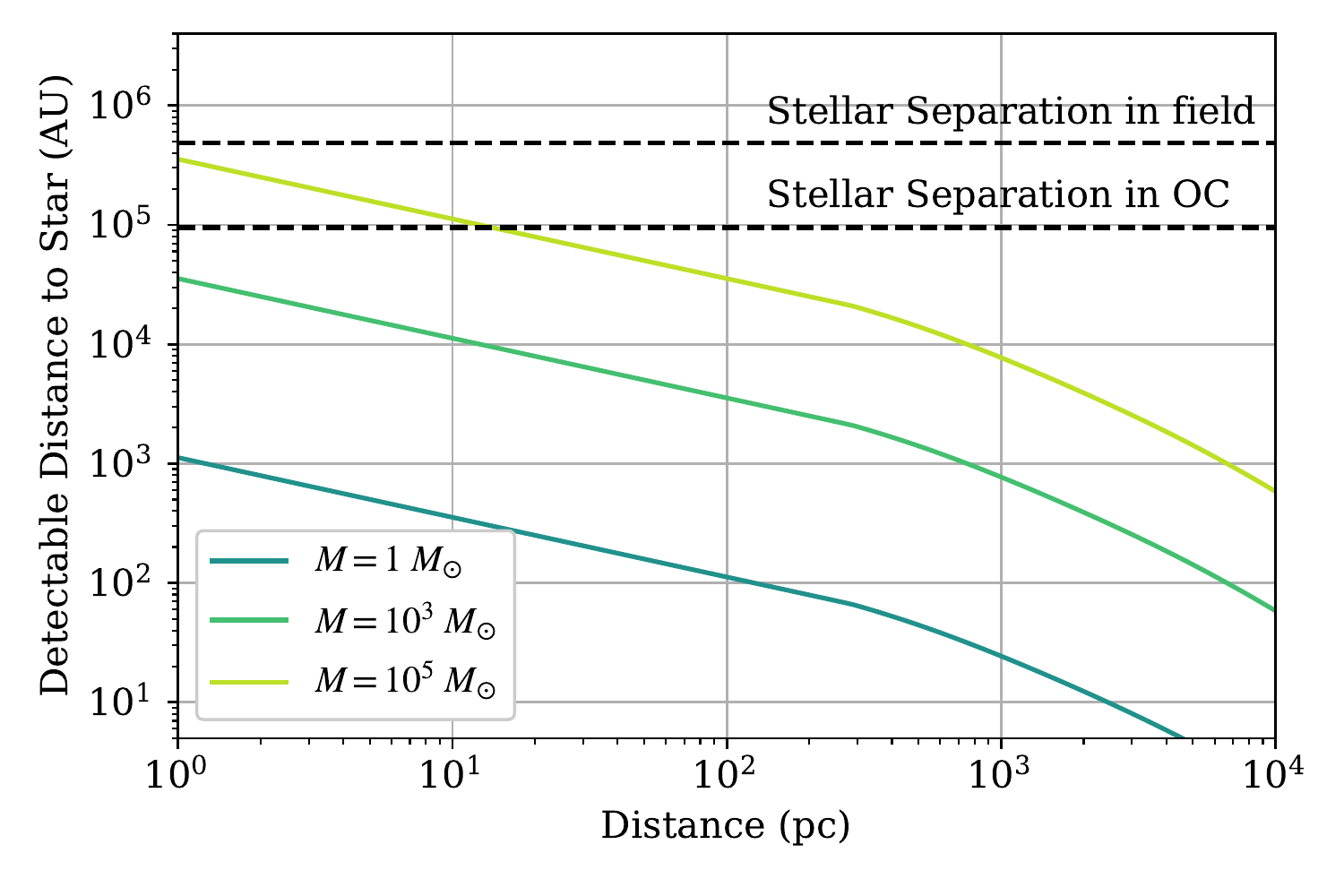}
    \caption{ To demonstrate that any individual accelerating signal is due to the fly-by of a hidden BH, the best method is to find correlated accelerations from multiple stars. This requires that multiple stars are sufficiently close to the hidden BH to produce a detectable acceleration. For the Milky Way field, only the most massive BHs ($\gtrsim10^5$~$\Msun$) are detectable and only within the nearest $
    \sim$10~pc. For open clusters (OC) with somewhat higher stellar densities, massive BHs are detectable to distances of $\sim$100~pc. }  
    \label{fig:hyper_accelerations}
    \end{center}
\end{figure}

Nevertheless, \gaia\ can rule out the existence of an IMBH in the Solar Neighborhood by searching for correlated accelerations detected by multiple stars. Since Gaia provides the acceleration vector projected on the sky, the intersection of two separate acceleration vectors indicates the position, distance, and mass of a dark object. A third star exhibiting acceleration overconstrains the system, allowing an important consistency check. The Solar Neighborhood has an average stellar separation of $\simeq$2.3 pc (the \gaia\ EDR3 catalog contains 2626 stars within the nearest 20 pc, leading to a local stellar density of $\simeq$0.08 pc$^{-3}$). Figure~\ref{fig:hyper_accelerations} shows the separation limit that the acceleration from an IMBH could be detectable by \gaia\ as a function of the BH mass and distance from the Sun. For multiple stars to exhibit accelerations from the same IMBH, the limiting stellar separation needs to be larger than the local stellar separation (indicated by the top horizontal, black, dashed line). In the Solar Neighborhood, the stellar separations are sufficiently large that even $10^5$~\Msun\ BHs are unlikely to be detectable. Open clusters with stellar densities reaching 10 pc$^{-3}$ offer a better opportunity to detect the existence of IMBHs. Figure~\ref{fig:hyper_accelerations} shows that IMBHs with masses of $\simeq10^5$~\Msun\ could be detectable in open clusters through astrometric acceleration. However, in their most extreme simulations of runaway collisions in the high density regions of young, stellar clusters, \citet{di_carlo2021} find IMBHs with masses no greater than a few $10^4$~\Msun\ and typically have much lower masses of $\sim10^2$~\Msun. We consider the prospects for detection dim, but nevertheless worthy of investigation.

\section{Discussion and Conclusions}
\label{sec:conclusions}

The third data release from \gaia\ contains $\simeq3\times10^5$ separate stars with detected astrometric accelerations. While astrometric accelerations can be caused by compact object (paper II), and stellar and substellar \citep[e.g.,][]{brandt2019} companions, in this work we explore the possibility that some fraction of these accelerations could be caused by hyperbolic fly-bys of dark, massive objects. Using a realistic prediction for \gaia's end-of-mission detection sensitivity to astrometric acceleration and under the extreme assumption that all of the Milky Way's dark matter is due to MACHOs, we find that none of the $\simeq3\times10^5$ stars exhibiting accelerations are likely to be due to fly-by interactions. Our conclusion is robust to even large inaccuracies in our assumptions about the density and velocity of putative MACHOs. Furthermore, even in the unlikely event that an unbound object induces a detectable acceleration, we are unaware of any unique observational signature that can separate an astrometric fly-by from the acceleration induced by a bound object on a wide orbit.

In principle, \gaia\ is also sensitive to the detection of rogue planets and planetesimals \citep{lissauer1987} traversing near a luminous star; a Jupiter-mass planet would induce a detectable acceleration on a star if it goes within $\sim2$~a.u. However, unbound planets likely traverse too quickly for \gaia\ to efficiently detect them. Furthermore, Figure~\ref{fig:N_detections} shows a clear decrease in detection probability for lower mass fly-bys; even if all the dark matter in the Milky Way is comprised of planet-mass objects, we again find that \gaia\ is unlikely to detect any of them. Our conclusions therefore also apply to the possibility of fly-by trajectories of luminous stars, free-floating planets, and asteroids.

Likewise we can consider the possibility of astrometric accelerations induced by two luminous stars traversing close to one another as they orbit around the Milky Way. Since the interaction velocities are an order of magnitude smaller while the local stellar densities in the disk is an order of magnitude higher than the local dark matter density, the accelerations induced by luminous stars are somewhat higher. However, this is offset by the fact that \gaia\ detects the acceleration of the system's photocenter, which both stars contribute to. Such interactions are only detectable if there is a large difference in luminosity between the two components. A complete investigation of the detectability of such interactions is outside the scope of this work. Here we comment that the detection rate would need to be several orders of magnitude larger to produce any meaningful fraction of the $\simeq3\times10^5$ detections within the \gaia\ DR3 catalog.

Finally, we consider the possibility that the Milky Way potential itself produces a fraction of the detected accelerations. At a rotational velocity of 220 km s$^{-1}$ and a distance from the Milky Way center of 8.3 kpc, the Sun feels an acceleration of $\sim10^{-8}$ cm s$^{-2}$, far too low to be detectable by even the nearby, best-measured stars with \gaia's end-of-mission precision. Only by correlating the astrometry of the entire \gaia\ catalog can such acceleration have the possibility of being detected \citep{buschmann2021}. Milky Way substructure, on the other hand, could induce a detectable signature if it is sufficiently clumpy. Further study is required to fully explore this possibility, but Equation~\ref{eq:accel_IMBH} provides some guidance. Using $10^{-4}$ cm s$^{-2}$ as a limiting detectable acceleration, a dark matter overdensity of mass $10^8$~\Msun\ would need to be confined within a $\sim10$~pc$^3$ volume to be detectable through correlated stellar accelerations. Pioneering studies in this direction have used the acceleration detected by pulsars \citep{bovy2020, chakrabarti2021, phillips2021}, but as yet, this possibility remains relatively unexplored for \gaia\ astrometry. Searching for stars with correlated accelerations in the \gaia\ catalog could help reveal any Milky Way substructure, if it exists with sufficient overdensity.

Having considered several alternative possibilities for the origin of astrometric accelerations exhibited by stars in the \gaia\ catalog, we conclude that the vast majority of them must be due to the existence of hidden compact object, stellar, and substellar companions in bound orbits. The current \gaia\ DR3 catalog contains $\simeq3\times10^5$ individual detections, far more than can be reasonably followed-up with observational resources. However, judicious follow-up may yield new, invaluable examples of gravitationally bound systems, something already being realized with small samples of nearby, bright stars exhibiting acceleration \citep[e.g.,][]{kuzuhara2022}. 

\acknowledgements

The author thanks Vicky Kalogera and Carl Rodriguez for useful discussions. J.J.A.\ acknowledges support from CIERA and Northwestern University through a Postdoctoral Fellowship. This work was initiated and performed in part at the Aspen Center for Physics, which is supported by National Science Foundation grant PHY-1607611.

\software{{\tt astropy} \citep{astropy}, {\tt NumPy} \citep{numpy}, {\tt SciPy} \citep{scipy}, {\tt matplotlib} \citep{matplotlib}}

\bibliographystyle{aasjournal}
\bibliography{gaia}

\begin{thebibliography}{}
\expandafter\ifx\csname natexlab\endcsname\relax\def\natexlab#1{#1}\fi

\bibitem[{{Abbott} {et~al.}(2020){Abbott}, {Abbott}, {Abraham}, {Acernese},
  {Ackley}, {Adams}, {Adhikari}, {Adya}, {Affeldt}, {Agathos}, {Agatsuma},
  {Aggarwal}, {Aguiar}, {Aich}, {Aiello}, {Ain}, {Ajith}, {Akcay}, {Allen},
  {Allocca}, {Altin}, {Amato}, {Anand}, {Ananyeva}, {Anderson}, {Anderson},
  {Angelova}, {Ansoldi}, {Antier}, {Appert}, {Arai}, {Araya}, {Areeda},
  {Ar{\`e}ne}, {Arnaud}, {Aronson}, {Arun}, {Asali}, {Ascenzi}, {Ashton},
  {Aston}, {Astone}, {Aubin}, {Aufmuth}, {AultONeal}, {Austin}, {Avendano},
  {Babak}, {Bacon}, {Badaracco}, {Bader}, {Bae}, {Baer}, {LIGO Scientific
  Collaboration}, \& {Virgo Collaboration}}]{GW190521}
{Abbott}, R., {Abbott}, T.~D., {Abraham}, S., {et~al.} 2020, \prl, 125, 101102

\bibitem[{{Alcock} {et~al.}(1993){Alcock}, {Akerlof}, {Allsman}, {Axelrod},
  {Bennett}, {Chan}, {Cook}, {Freeman}, {Griest}, {Marshall}, {Park},
  {Perlmutter}, {Peterson}, {Pratt}, {Quinn}, {Rodgers}, {Stubbs}, \&
  {Sutherland}}]{alcock1993}
{Alcock}, C., {Akerlof}, C.~W., {Allsman}, R.~A., {et~al.} 1993, \nat, 365, 621

\bibitem[{{Andrews} {et~al.}(2019){Andrews}, {Breivik}, \&
  {Chatterjee}}]{andrews2019}
{Andrews}, J.~J., {Breivik}, K., \& {Chatterjee}, S. 2019, \apj, 886, 68

\bibitem[{{Andrews} {et~al.}(2021){Andrews}, {Breivik}, {Chawla}, {Rodriguez},
  \& {Chatterjee}}]{andrews2021}
{Andrews}, J.~J., {Breivik}, K., {Chawla}, C., {Rodriguez}, C., \&
  {Chatterjee}, S. 2021, arXiv e-prints, arXiv:2110.05549

\bibitem[{{Andrews} \& {Kalogera}(2022)}]{andrews2022}
{Andrews}, J.~J., \& {Kalogera}, V. 2022, \apj, 930, 159

\bibitem[{{Astropy Collaboration} {et~al.}(2018){Astropy Collaboration},
  {Price-Whelan}, {Sip{\H{o}}cz}, {G{\"u}nther}, {Lim}, {Crawford}, {Conseil},
  {Shupe}, {Craig}, {Dencheva}, {Ginsburg}, {Vand erPlas}, {Bradley},
  {P{\'e}rez-Su{\'a}rez}, {de Val-Borro}, {Aldcroft}, {Cruz}, {Robitaille},
  {Tollerud}, {Ardelean}, {Babej}, {Bach}, {Bachetti}, {Bakanov}, {Bamford},
  {Barentsen}, {Barmby}, {Baumbach}, {Berry}, {Biscani}, {Boquien}, {Bostroem},
  {Bouma}, {Brammer}, {Bray}, {Breytenbach}, {Buddelmeijer}, {Burke},
  {Calderone}, {Cano Rodr{\'\i}guez}, {Cara}, {Cardoso}, {Cheedella}, {Copin},
  {Corrales}, {Crichton}, {D'Avella}, {Deil}, {Depagne}, {Dietrich}, {Donath},
  {Droettboom}, {Earl}, {Erben}, {Fabbro}, {Ferreira}, {Finethy}, {Fox},
  {Garrison}, {Gibbons}, {Goldstein}, {Gommers}, {Greco}, {Greenfield},
  {Groener}, {Grollier}, {Hagen}, {Hirst}, {Homeier}, {Horton}, {Hosseinzadeh},
  {Hu}, {Hunkeler}, {Ivezi{\'c}}, {Jain}, {Jenness}, {Kanarek}, {Kendrew},
  {Kern}, {Kerzendorf}, {Khvalko}, {King}, {Kirkby}, {Kulkarni}, {Kumar},
  {Lee}, {Lenz}, {Littlefair}, {Ma}, {Macleod}, {Mastropietro}, {McCully},
  {Montagnac}, {Morris}, {Mueller}, {Mumford}, {Muna}, {Murphy}, {Nelson},
  {Nguyen}, {Ninan}, {N{\"o}the}, {Ogaz}, {Oh}, {Parejko}, {Parley}, {Pascual},
  {Patil}, {Patil}, {Plunkett}, {Prochaska}, {Rastogi}, {Reddy Janga},
  {Sabater}, {Sakurikar}, {Seifert}, {Sherbert}, {Sherwood-Taylor}, {Shih},
  {Sick}, {Silbiger}, {Singanamalla}, {Singer}, {Sladen}, {Sooley},
  {Sornarajah}, {Streicher}, {Teuben}, {Thomas}, {Tremblay}, {Turner},
  {Terr{\'o}n}, {van Kerkwijk}, {de la Vega}, {Watkins}, {Weaver}, {Whitmore},
  {Woillez}, {Zabalza}, \& {Astropy Contributors}}]{astropy}
{Astropy Collaboration}, {Price-Whelan}, A.~M., {Sip{\H{o}}cz}, B.~M., {et~al.}
  2018, \aj, 156, 123

\bibitem[{{Bahcall} {et~al.}(1985){Bahcall}, {Hut}, \& {Tremaine}}]{bahcall85}
{Bahcall}, J.~N., {Hut}, P., \& {Tremaine}, S. 1985, \apj, 290, 15

\bibitem[{{Barstow} {et~al.}(2014){Barstow}, {Casewell}, {Catalan},
  {Copperwheat}, {Gaensicke}, {Garcia-Berro}, {Hambly}, {Heber}, {Holberg},
  {Isern}, {Jeffery}, {Jordan}, {Lawrie}, {Lynas-Gray}, {Maccarone}, {Marsh},
  {Parsons}, {Silvotti}, {Subasavage}, {Torres}, \& {Wheatley}}]{Barstow2014}
{Barstow}, M.~A., {Casewell}, S.~L., {Catalan}, S., {et~al.} 2014, arXiv
  e-prints, arXiv:1407.6163

\bibitem[{{Bennett} {et~al.}(2002){Bennett}, {Becker}, {Quinn}, {Tomaney},
  {Alcock}, {Allsman}, {Alves}, {Axelrod}, {Calitz}, {Cook}, {Drake},
  {Fragile}, {Freeman}, {Geha}, {Griest}, {Johnson}, {Keller}, {Laws},
  {Lehner}, {Marshall}, {Minniti}, {Nelson}, {Peterson}, {Popowski}, {Pratt},
  {Quinn}, {Rhie}, {Stubbs}, {Sutherland}, {Vandehei}, {Welch}, {MACHO
  Collaboration}, \& {MPS Collaboration}}]{bennett2002}
{Bennett}, D.~P., {Becker}, A.~C., {Quinn}, J.~L., {et~al.} 2002, \apj, 579,
  639

\bibitem[{{Bennett} \& {Bovy}(2019)}]{bennett2019}
{Bennett}, M., \& {Bovy}, J. 2019, \mnras, 482, 1417

\bibitem[{{Biller} {et~al.}(2022){Biller}, {Grandjean}, {Messina}, {Desidera},
  {Delorme}, {Lagrange}, {Hambsch}, {Mesa}, {Janson}, {Gratton}, {D'Orazi},
  {Langlois}, {Maire}, {Schlieder}, {Henning}, {Zurlo}, {Hagelberg},
  {Brown-Sevilla}, {Romero}, {Bonnefoy}, {Chauvin}, {Feldt}, {Meyer}, {Vigan},
  {Pavlov}, {Soenke}, {LeMignant}, \& {Roux}}]{biller2022}
{Biller}, B.~A., {Grandjean}, A., {Messina}, S., {et~al.} 2022, \aap, 658, A145

\bibitem[{{Blaineau} {et~al.}(2022){Blaineau}, {Moniez}, {Afonso}, {Albert},
  {Ansari}, {Aubourg}, {Coutures}, {Glicenstein}, {Goldman}, {Hamadache},
  {Lasserre}, {LeGuillou}, {Lesquoy}, {Magneville}, {Marquette},
  {Palanque-Delabrouille}, {Perdereau}, {Rich}, {Spiro}, \&
  {Tisserand}}]{blaineau2022}
{Blaineau}, T., {Moniez}, M., {Afonso}, C., {et~al.} 2022, arXiv e-prints,
  arXiv:2202.13819

\bibitem[{{Bonavita} {et~al.}(2022){Bonavita}, {Fontanive}, {Gratton}, {Muzic},
  {Desidera}, {Biller}, {Scholz}, {Sozzetti}, \& {Squicciarini}}]{bonavita2022}
{Bonavita}, M., {Fontanive}, C., {Gratton}, R., {et~al.} 2022, arXiv e-prints,
  arXiv:2205.02213

\bibitem[{{Bovy}(2020)}]{bovy2020}
{Bovy}, J. 2020, arXiv e-prints, arXiv:2012.02169

\bibitem[{{Bowler} {et~al.}(2021){Bowler}, {Endl}, {Cochran}, {MacQueen},
  {Crepp}, {Doppmann}, {Dulz}, {Brandt}, {Mirek Brandt}, {Li}, {Dupuy},
  {Franson}, {Kratter}, {Morley}, \& {Zhou}}]{bowler2021}
{Bowler}, B.~P., {Endl}, M., {Cochran}, W.~D., {et~al.} 2021, \apjl, 913, L26

\bibitem[{{Brandt} {et~al.}(2021){Brandt}, {Dupuy}, {Li}, {Chen}, {Brandt},
  {Wong}, {Currie}, {Bowler}, {Liu}, {Best}, \& {Phillips}}]{brandt2021b}
{Brandt}, G.~M., {Dupuy}, T.~J., {Li}, Y., {et~al.} 2021, \aj, 162, 301

\bibitem[{{Brandt}(2018)}]{Brandt2018}
{Brandt}, T.~D. 2018, \apjs, 239, 31

\bibitem[{{Brandt}(2021)}]{brandt2021a}
---. 2021, \apjs, 254, 42

\bibitem[{{Brandt} {et~al.}(2019){Brandt}, {Dupuy}, \& {Bowler}}]{brandt2019}
{Brandt}, T.~D., {Dupuy}, T.~J., \& {Bowler}, B.~P. 2019, \aj, 158, 140

\bibitem[{{Breivik} {et~al.}(2019){Breivik}, {Chatterjee}, \&
  {Andrews}}]{Breivik2018}
{Breivik}, K., {Chatterjee}, S., \& {Andrews}, J.~J. 2019, \apj, 878, L4

\bibitem[{{Breivik} {et~al.}(2017){Breivik}, {Chatterjee}, \&
  {Larson}}]{Breivik2017}
{Breivik}, K., {Chatterjee}, S., \& {Larson}, S.~L. 2017, \apjl, 850, L13

\bibitem[{{Brown}(2021)}]{brown2021}
{Brown}, A. G.~A. 2021, \araa, 59, arXiv:2102.11712

\bibitem[{{Buschmann} {et~al.}(2021){Buschmann}, {Safdi}, \&
  {Schutz}}]{buschmann2021}
{Buschmann}, M., {Safdi}, B.~R., \& {Schutz}, K. 2021, \prl, 127, 241104

\bibitem[{Chakrabarti {et~al.}(2021)Chakrabarti, Chang, Lam, Vigeland, \&
  Quillen}]{chakrabarti2021}
Chakrabarti, S., Chang, P., Lam, M.~T., Vigeland, S.~J., \& Quillen, A.~C.
  2021, The Astrophysical Journal Letters, 907, L26

\bibitem[{Chakrabarti {et~al.}(2020)Chakrabarti, Wright, Chang, Quillen, Craig,
  Territo, D'Onghia, Johnston, Rosa, Huber, Rhode, \&
  Nielsen}]{chakrabarti2020}
Chakrabarti, S., Wright, J., Chang, P., {et~al.} 2020, The Astrophysical
  Journal, 902, L28

\bibitem[{{Chawla} {et~al.}(2021){Chawla}, {Chatterjee}, {Breivik}, {Krishna
  Moorthy}, {Andrews}, \& {Sanderson}}]{chawla2021}
{Chawla}, C., {Chatterjee}, S., {Breivik}, K., {et~al.} 2021, arXiv e-prints,
  arXiv:2110.05979

\bibitem[{{Davis} {et~al.}(2011){Davis}, {Narayan}, {Zhu}, {Barret}, {Farrell},
  {Godet}, {Servillat}, \& {Webb}}]{davis2011}
{Davis}, S.~W., {Narayan}, R., {Zhu}, Y., {et~al.} 2011, \apj, 734, 111

\bibitem[{{Di Carlo} {et~al.}(2021){Di Carlo}, {Mapelli}, {Pasquato},
  {Rastello}, {Ballone}, {Dall'Amico}, {Giacobbo}, {Iorio}, {Spera},
  {Torniamenti}, \& {Haardt}}]{di_carlo2021}
{Di Carlo}, U.~N., {Mapelli}, M., {Pasquato}, M., {et~al.} 2021, \mnras, 507,
  5132

\bibitem[{{Dupuy} {et~al.}(2022){Dupuy}, {Brandt}, \& {Brandt}}]{dupuy2022}
{Dupuy}, T.~J., {Brandt}, G.~M., \& {Brandt}, T.~D. 2022, \mnras, 509, 4411

\bibitem[{{Errico} {et~al.}(2022){Errico}, {Wittenmyer}, {Horner}, {Li},
  {Brandt}, {Kane}, {Fetherolf}, {Holt}, {Carter}, {Butler}, {Tinney},
  {Ballard}, {Bowler}, {Kielkopf}, {Liu}, {Plavchan}, {Shporer}, {Zhang},
  {Wright}, {Addison}, {Mengel}, \& {Okumura}}]{errico2022}
{Errico}, A., {Wittenmyer}, R.~A., {Horner}, J., {et~al.} 2022, arXiv e-prints,
  arXiv:2204.05711

\bibitem[{{Feng} {et~al.}(2021){Feng}, {Butler}, {Jones}, {Phillips}, {Vogt},
  {Oppenheimer}, {Holden}, {Burt}, \& {Boss}}]{feng2021}
{Feng}, F., {Butler}, R.~P., {Jones}, H. R.~A., {et~al.} 2021, \mnras, 507,
  2856

\bibitem[{{Franco} {et~al.}(2021){Franco}, {Nucita}, {De Paolis}, {Strafella},
  \& {Maiorano}}]{franco2021}
{Franco}, A., {Nucita}, A.~A., {De Paolis}, F., {Strafella}, F., \& {Maiorano},
  M. 2021, arXiv e-prints, arXiv:2110.11047

\bibitem[{{Franson} {et~al.}(2022){Franson}, {Bowler}, {Brandt}, {Dupuy},
  {Tran}, {Brandt}, {Li}, \& {Kraus}}]{franson2022}
{Franson}, K., {Bowler}, B.~P., {Brandt}, T.~D., {et~al.} 2022, \aj, 163, 50

\bibitem[{{Gaia Collaboration} {et~al.}(2016){Gaia Collaboration}, {Prusti},
  {de Bruijne}, {Brown}, {Vallenari}, {Babusiaux}, {Bailer-Jones}, {Bastian},
  {Biermann}, {Evans}, \& et~al.}]{Gaia_mission}
{Gaia Collaboration}, {Prusti}, T., {de Bruijne}, J.~H.~J., {et~al.} 2016,
  \aap, 595, A1

\bibitem[{{Gaia Collaboration} {et~al.}(2021){Gaia Collaboration}, {Brown},
  {Vallenari}, {Prusti}, {de Bruijne}, {Babusiaux}, {Biermann}, {Creevey},
  {Evans}, {Eyer}, {Hutton}, {Jansen}, {Jordi}, {Klioner}, {Lammers},
  {Lindegren}, {Luri}, {Mignard}, {Panem}, {Pourbaix}, {Randich}, {Sartoretti},
  {Soubiran}, {Walton}, {Arenou}, {Bailer-Jones}, {Bastian}, {Cropper},
  {Drimmel}, {Katz}, {Lattanzi}, {van Leeuwen}, {Bakker}, {Cacciari},
  {Casta{\~n}eda}, {De Angeli}, {Ducourant}, {Fabricius}, {Fouesneau},
  {Fr{\'e}mat}, {Guerra}, {Guerrier}, {Guiraud}, {Jean-Antoine Piccolo},
  {Masana}, {Messineo}, {Mowlavi}, {Nicolas}, {Nienartowicz}, {Pailler},
  {Panuzzo}, {Riclet}, {Roux}, {Seabroke}, {Sordo}, {Tanga}, {Th{\'e}venin},
  {Gracia-Abril}, {Portell}, {Teyssier}, {Altmann}, {Andrae}, {Bellas-Velidis},
  {Benson}, {Berthier}, {Blomme}, {Brugaletta}, {Burgess}, {Busso}, {Carry},
  {Cellino}, {Cheek}, {Clementini}, {Damerdji}, {Davidson}, {Delchambre},
  {Dell'Oro}, {Fern{\'a}ndez-Hern{\'a}ndez}, {Galluccio}, {Garc{\'\i}a-Lario},
  {Garcia-Reinaldos}, {Gonz{\'a}lez-N{\'u}{\~n}ez}, {Gosset}, {Haigron},
  {Halbwachs}, {Hambly}, {Harrison}, {Hatzidimitriou}, {Heiter},
  {Hern{\'a}ndez}, {Hestroffer}, {Hodgkin}, {Holl}, {Jan{\ss}en}, {Jevardat de
  Fombelle}, {Jordan}, {Krone-Martins}, {Lanzafame}, {L{\"o}ffler}, {Lorca},
  {Manteiga}, {Marchal}, {Marrese}, {Moitinho}, {Mora}, {Muinonen}, {Osborne},
  {Pancino}, {Pauwels}, {Petit}, {Recio-Blanco}, {Richards}, {Riello},
  {Rimoldini}, {Robin}, {Roegiers}, {Rybizki}, {Sarro}, {Siopis}, {Smith},
  {Sozzetti}, {Ulla}, {Utrilla}, {van Leeuwen}, {van Reeven}, {Abbas}, {Abreu
  Aramburu}, {Accart}, {Aerts}, {Aguado}, {Ajaj}, {Altavilla}, {{\'A}lvarez},
  {{\'A}lvarez Cid-Fuentes}, {Alves}, {Anderson}, {Anglada Varela}, {Antoja},
  {Audard}, {Baines}, {Baker}, {Balaguer-N{\'u}{\~n}ez}, {Balbinot}, {Balog},
  {Barache}, {Barbato}, {Barros}, {Barstow}, {Bartolom{\'e}}, {Bassilana},
  {Bauchet}, {Baudesson-Stella}, {Becciani}, {Bellazzini}, {Bernet}, {Bertone},
  {Bianchi}, {Blanco-Cuaresma}, {Boch}, {Bombrun}, {Bossini}, {Bouquillon},
  {Bragaglia}, {Bramante}, {Breedt}, {Bressan}, {Brouillet}, {Bucciarelli},
  {Burlacu}, {Busonero}, {Butkevich}, {Buzzi}, {Caffau}, {Cancelliere},
  {C{\'a}novas}, {Cantat-Gaudin}, {Carballo}, {Carlucci}, {Carnerero},
  {Carrasco}, {Casamiquela}, {Castellani}, {Castro-Ginard}, {Castro Sampol},
  {Chaoul}, {Charlot}, {Chemin}, {Chiavassa}, {Cioni}, {Comoretto}, {Cooper},
  {Cornez}, {Cowell}, {Crifo}, {Crosta}, {Crowley}, {Dafonte}, {Dapergolas},
  {David}, {David}, {de Laverny}, {De Luise}, {De March}, {De Ridder}, {de
  Souza}, {de Teodoro}, {de Torres}, {del Peloso}, {del Pozo}, {Delbo},
  {Delgado}, {Delgado}, {Delisle}, {Di Matteo}, {Diakite}, {Diener},
  {Distefano}, {Dolding}, {Eappachen}, {Edvardsson}, {Enke}, {Esquej}, {Fabre},
  {Fabrizio}, {Faigler}, {Fedorets}, {Fernique}, {Fienga}, {Figueras},
  {Fouron}, {Fragkoudi}, {Fraile}, {Franke}, {Gai}, {Garabato},
  {Garcia-Gutierrez}, {Garc{\'\i}a-Torres}, {Garofalo}, {Gavras}, {Gerlach},
  {Geyer}, {Giacobbe}, {Gilmore}, {Girona}, {Giuffrida}, {Gomel}, {Gomez},
  {Gonzalez-Santamaria}, {Gonz{\'a}lez-Vidal}, {Granvik},
  {Guti{\'e}rrez-S{\'a}nchez}, {Guy}, {Hauser}, {Haywood}, {Helmi}, {Hidalgo},
  {Hilger}, {H{\l}adczuk}, {Hobbs}, {Holland}, {Huckle}, {Jasniewicz},
  {Jonker}, {Juaristi Campillo}, {Julbe}, {Karbevska}, {Kervella}, {Khanna},
  {Kochoska}, {Kontizas}, {Kordopatis}, {Korn}, {Kostrzewa-Rutkowska},
  {Kruszy{\'n}ska}, {Lambert}, {Lanza}, {Lasne}, {Le Campion}, {Le Fustec},
  {Lebreton}, {Lebzelter}, {Leccia}, {Leclerc}, {Lecoeur-Taibi}, {Liao},
  {Licata}, {Lindstr{\o}m}, {Lister}, {Livanou}, {Lobel}, {Madrero Pardo},
  {Managau}, {Mann}, {Marchant}, {Marconi}, {Marcos Santos}, {Marinoni},
  {Marocco}, {Marshall}, {Martin Polo}, {Mart{\'\i}n-Fleitas}, {Masip},
  {Massari}, {Mastrobuono-Battisti}, {Mazeh}, {McMillan}, {Messina},
  {Michalik}, {Millar}, {Mints}, {Molina}, {Molinaro}, {Moln{\'a}r},
  {Montegriffo}, {Mor}, {Morbidelli}, {Morel}, {Morris}, {Mulone}, {Munoz},
  {Muraveva}, {Murphy}, {Musella}, {Noval}, {Ord{\'e}novic}, {Orr{\`u}},
  {Osinde}, {Pagani}, {Pagano}, {Palaversa}, {Palicio}, {Panahi}, {Pawlak},
  {Pe{\~n}alosa Esteller}, {Penttil{\"a}}, {Piersimoni}, {Pineau}, {Plachy},
  {Plum}, {Poggio}, {Poretti}, {Poujoulet}, {Pr{\v{s}}a}, {Pulone}, {Racero},
  {Ragaini}, {Rainer}, {Raiteri}, {Rambaux}, {Ramos}, {Ramos-Lerate}, {Re
  Fiorentin}, {Regibo}, {Reyl{\'e}}, {Ripepi}, {Riva}, {Rixon}, {Robichon},
  {Robin}, {Roelens}, {Rohrbasser}, {Romero-G{\'o}mez}, {Rowell}, {Royer},
  {Rybicki}, {Sadowski}, {Sagrist{\`a} Sell{\'e}s}, {Sahlmann}, {Salgado},
  {Salguero}, {Samaras}, {Sanchez Gimenez}, {Sanna}, {Santove{\~n}a},
  {Sarasso}, {Schultheis}, {Sciacca}, {Segol}, {Segovia}, {S{\'e}gransan},
  {Semeux}, {Shahaf}, {Siddiqui}, {Siebert}, {Siltala}, {Slezak}, {Smart},
  {Solano}, {Solitro}, {Souami}, {Souchay}, {Spagna}, {Spoto}, {Steele},
  {Steidelm{\"u}ller}, {Stephenson}, {S{\"u}veges}, {Szabados}, {Szegedi-Elek},
  {Taris}, {Tauran}, {Taylor}, {Teixeira}, {Thuillot}, {Tonello}, {Torra},
  {Torra}, {Turon}, {Unger}, {Vaillant}, {van Dillen}, {Vanel}, {Vecchiato},
  {Viala}, {Vicente}, {Voutsinas}, {Weiler}, {Wevers}, {Wyrzykowski}, {Yoldas},
  {Yvard}, {Zhao}, {Zorec}, {Zucker}, {Zurbach}, \& {Zwitter}}]{gaia_EDR3}
{Gaia Collaboration}, {Brown}, A.~G.~A., {Vallenari}, A., {et~al.} 2021, \aap,
  649, A1

\bibitem[{{Gais} {et~al.}(2022){Gais}, {Ng}, {Seo}, {Wong}, \& {Li}}]{gais2022}
{Gais}, J., {Ng}, K., {Seo}, E., {Wong}, K. W.~K., \& {Li}, T. G.~F. 2022,
  arXiv e-prints, arXiv:2201.01817

\bibitem[{{Golovich} {et~al.}(2022){Golovich}, {Dawson}, {Bartoli{\'c}}, {Lam},
  {Lu}, {Medford}, {Schneider}, {Chapline}, {Schlafly}, {Drlica-Wagner}, \&
  {Pruett}}]{golovich2022}
{Golovich}, N., {Dawson}, W., {Bartoli{\'c}}, F., {et~al.} 2022, \apjs, 260, 2

\bibitem[{{Gould} \& {Salim}(2002)}]{gould2002}
{Gould}, A., \& {Salim}, S. 2002, \apj, 572, 944

\bibitem[{Harris {et~al.}(2020)Harris, Millman, van~der Walt, Gommers,
  Virtanen, Cournapeau, Wieser, Taylor, Berg, Smith, Kern, Picus, Hoyer, van
  Kerkwijk, Brett, Haldane, del R{\'{i}}o, Wiebe, Peterson,
  G{\'{e}}rard-Marchant, Sheppard, Reddy, Weckesser, Abbasi, Gohlke, \&
  Oliphant}]{numpy}
Harris, C.~R., Millman, K.~J., van~der Walt, S.~J., {et~al.} 2020, Nature, 585,
  357

\bibitem[{{Hartkopf} {et~al.}(1996){Hartkopf}, {Mason}, \&
  {McAlister}}]{Hartkopf1996}
{Hartkopf}, W.~I., {Mason}, B.~D., \& {McAlister}, H.~A. 1996, \aj, 111, 370

\bibitem[{Hunter(2007)}]{matplotlib}
Hunter, J.~D. 2007, Computing in Science Engineering, 9, 90

\bibitem[{{Kains} {et~al.}(2018){Kains}, {Calamida}, {Sahu}, {Anderson},
  {Casertano}, \& {Bramich}}]{kains2018}
{Kains}, N., {Calamida}, A., {Sahu}, K.~C., {et~al.} 2018, \apj, 867, 37

\bibitem[{{Kervella} {et~al.}(2019){Kervella}, {Arenou}, {Mignard}, \&
  {Th{\'e}venin}}]{kervella2019a}
{Kervella}, P., {Arenou}, F., {Mignard}, F., \& {Th{\'e}venin}, F. 2019, \aap,
  623, A72

\bibitem[{{Kounkel} \& {Covey}(2019)}]{kounkel2019}
{Kounkel}, M., \& {Covey}, K. 2019, \aj, 158, 122

\bibitem[{{Kuzuhara} {et~al.}(2022){Kuzuhara}, {Currie}, {Takarada}, {Brandt},
  {Sato}, {Uyama}, {Janson}, {Chilcote}, {Tobin}, {Lawson}, {Hori}, {Guyon},
  {Groff}, {Lozi}, {Vievard}, {Sahoo}, {Deo}, {Jovanovic}, {Ahn}, {Martinache},
  {Skaf}, {Akiyama}, {Norris}, {Bonnefoy}, {He{\l}miniak}, {Kudo}, {McElwain},
  {Samland}, {Wagner}, {Wisniewski}, {Knapp}, {Kwon}, {Nishikawa}, {Serabyn},
  {Hayashi}, \& {Tamura}}]{kuzuhara2022}
{Kuzuhara}, M., {Currie}, T., {Takarada}, T., {et~al.} 2022, arXiv e-prints,
  arXiv:2205.02729

\bibitem[{{Lai} {et~al.}(2018){Lai}, {Hannuksela}, {Herrera-Mart{\'\i}n},
  {Diego}, {Broadhurst}, \& {Li}}]{lai2018}
{Lai}, K.-H., {Hannuksela}, O.~A., {Herrera-Mart{\'\i}n}, A., {et~al.} 2018,
  \prd, 98, 083005

\bibitem[{{Lam} {et~al.}(2022){Lam}, {Lu}, {Udalski}, {Bond}, {Bennett},
  {Skowron}, {Mroz}, {Poleski}, {Sumi}, {Szymanski}, {Kozlowski},
  {Pietrukowicz}, {Soszynski}, {Ulaczyk}, {Wyrzykowski}, {Miyazaki}, {Suzuki},
  {Koshimoto}, {Rattenbury}, {Hosek}, {Abe}, {Barry}, {Bhattacharya}, {Fukui},
  {Fujii}, {Hirao}, {Itow}, {Kirikawa}, {Kondo}, {Matsubara}, {Matsumoto},
  {Muraki}, {Olmschenk}, {Ranc}, {Okamura}, {Satoh}, {Ishitani Silva}, {Toda},
  {Tristram}, {Vandorou}, {Yama}, {Abrams}, {Agarwal}, {Rose}, \&
  {Terry}}]{lam2022}
{Lam}, C.~Y., {Lu}, J.~R., {Udalski}, A., {et~al.} 2022, arXiv e-prints,
  arXiv:2202.01903

\bibitem[{{Li} {et~al.}(2021){Li}, {Brandt}, {Brandt}, {Dupuy}, {Michalik},
  {Jensen-Clem}, {Zeng}, {Faherty}, \& {Mitra}}]{li2021}
{Li}, Y., {Brandt}, T.~D., {Brandt}, G.~M., {et~al.} 2021, \aj, 162, 266

\bibitem[{{Lindegren} {et~al.}(2021){Lindegren}, {Klioner}, {Hern{\'a}ndez},
  {Bombrun}, {Ramos-Lerate}, {Steidelm{\"u}ller}, {Bastian}, {Biermann}, {de
  Torres}, {Gerlach}, {Geyer}, {Hilger}, {Hobbs}, {Lammers}, {McMillan},
  {Stephenson}, {Casta{\~n}eda}, {Davidson}, {Fabricius}, {Gracia-Abril},
  {Portell}, {Rowell}, {Teyssier}, {Torra}, {Bartolom{\'e}}, {Clotet},
  {Garralda}, {Gonz{\'a}lez-Vidal}, {Torra}, {Abbas}, {Altmann}, {Anglada
  Varela}, {Balaguer-N{\'u}{\~n}ez}, {Balog}, {Barache}, {Becciani}, {Bernet},
  {Bertone}, {Bianchi}, {Bouquillon}, {Brown}, {Bucciarelli}, {Busonero},
  {Butkevich}, {Buzzi}, {Cancelliere}, {Carlucci}, {Charlot}, {Cioni},
  {Crosta}, {Crowley}, {del Peloso}, {del Pozo}, {Drimmel}, {Esquej}, {Fienga},
  {Fraile}, {Gai}, {Garcia-Reinaldos}, {Guerra}, {Hambly}, {Hauser},
  {Jan{\ss}en}, {Jordan}, {Kostrzewa-Rutkowska}, {Lattanzi}, {Liao}, {Licata},
  {Lister}, {L{\"o}ffler}, {Marchant}, {Masip}, {Mignard}, {Mints}, {Molina},
  {Mora}, {Morbidelli}, {Murphy}, {Pagani}, {Panuzzo}, {Pe{\~n}alosa Esteller},
  {Poggio}, {Re Fiorentin}, {Riva}, {Sagrist{\`a} Sell{\'e}s}, {Sanchez
  Gimenez}, {Sarasso}, {Sciacca}, {Siddiqui}, {Smart}, {Souami}, {Spagna},
  {Steele}, {Taris}, {Utrilla}, {van Reeven}, \& {Vecchiato}}]{lindegren2021}
{Lindegren}, L., {Klioner}, S.~A., {Hern{\'a}ndez}, J., {et~al.} 2021, \aap,
  649, A2

\bibitem[{{Lissauer}(1987)}]{lissauer1987}
{Lissauer}, J.~J. 1987, \icarus, 69, 249

\bibitem[{{Mao} {et~al.}(2002){Mao}, {Smith}, {Wo{\'z}niak}, {Udalski},
  {Szyma{\'n}ski}, {Kubiak}, {Pietrzy{\'n}ski}, {Soszy{\'n}ski}, \&
  {{\.Z}ebru{\'n}}}]{mao2002}
{Mao}, S., {Smith}, M.~C., {Wo{\'z}niak}, P., {et~al.} 2002, \mnras, 329, 349

\bibitem[{{Mashian} \& {Loeb}(2017)}]{Mashian2017}
{Mashian}, N., \& {Loeb}, A. 2017, \mnras, 470, 2611

\bibitem[{{Mason} {et~al.}(1999){Mason}, {Douglass}, \& {Hartkopf}}]{Mason1999}
{Mason}, B.~D., {Douglass}, G.~G., \& {Hartkopf}, W.~I. 1999, \aj, 117, 1023

\bibitem[{{Mirhosseini} \& {Moniez}(2018)}]{mirhosseini2018}
{Mirhosseini}, A., \& {Moniez}, M. 2018, \aap, 618, L4

\bibitem[{{Paczynski}(1986)}]{paczynski86}
{Paczynski}, B. 1986, \apj, 304, 1

\bibitem[{{Paynter} {et~al.}(2021){Paynter}, {Webster}, \&
  {Thrane}}]{paynter2021}
{Paynter}, J., {Webster}, R., \& {Thrane}, E. 2021, Nature Astronomy, 5, 560

\bibitem[{{Penoyre} {et~al.}(2020){Penoyre}, {Belokurov}, {Wyn Evans},
  {Everall}, \& {Koposov}}]{penoyre2020}
{Penoyre}, Z., {Belokurov}, V., {Wyn Evans}, N., {Everall}, A., \& {Koposov},
  S.~E. 2020, \mnras, 495, 321

\bibitem[{{Perryman} {et~al.}(2014){Perryman}, {Hartman}, {Bakos}, \&
  {Lindegren}}]{Perryman2014}
{Perryman}, M., {Hartman}, J., {Bakos}, G.~{\'A}., \& {Lindegren}, L. 2014,
  \apj, 797, 14

\bibitem[{Phillips {et~al.}(2021)Phillips, Ravi, Ebadi, \&
  Walsworth}]{phillips2021}
Phillips, D.~F., Ravi, A., Ebadi, R., \& Walsworth, R.~L. 2021, Phys. Rev.
  Lett., 126, 141103

\bibitem[{{Pourbaix} \& {Jorissen}(2000)}]{Pourbaix2000}
{Pourbaix}, D., \& {Jorissen}, A. 2000, \aaps, 145, 161

\bibitem[{Ravi {et~al.}(2019)Ravi, Langellier, Phillips, Buschmann, Safdi, \&
  Walsworth}]{ravi2019}
Ravi, A., Langellier, N., Phillips, D.~F., {et~al.} 2019, Phys. Rev. Lett.,
  123, 091101

\bibitem[{{Read}(2014)}]{read2014}
{Read}, J.~I. 2014, Journal of Physics G Nuclear Physics, 41, 063101

\bibitem[{{Sahu} {et~al.}(2022){Sahu}, {Anderson}, {Casertano}, {Bond},
  {Udalski}, {Dominik}, {Calamida}, {Bellini}, {Brown}, {Rejkuba}, {Bajaj},
  {Kains}, {Ferguson}, {Fryer}, {Yock}, {Mroz}, {Kozlowski}, {Pietrukowicz},
  {Poleski}, {Skowron}, {Soszynski}, {Szymanski}, {Ulaczyk}, {Wyrzykowski},
  {Beaulieu}, {Marquette}, {Cole}, {Hill}, {Dieters}, {Coutures},
  {Dominis-Prester}, {Bachelet}, {Menzies}, {Albrow}, {Pollard}, {Gould},
  {Yee}, {Allen}, {de Almeida}, {Christie}, {Drummond}, {Gal-Yam}, {Gorbikov},
  {Jablonski}, {Lee}, {Maoz}, {Manulis}, {McCormick}, {Natusch}, {Pogge},
  {Shvartzvald}, {Jorgensen}, {Alsubai}, {Andersen}, {Bozza}, {Calchi Novati},
  {Hinse}, {Hundertmark}, {Husser}, {Kerins}, {Longa-Pena}, {Mancini}, {Penny},
  {Rahvar}, {Ricci}, {Sajadian}, {Skottfelt}, {Snodgrass}, {Southworth},
  {Tregloan-Reed}, {Wambsganss}, {Wertz}, {Tsapras}, {Street}, {Bramich},
  {Horne}, \& {Steele}}]{sahu2022}
{Sahu}, K.~C., {Anderson}, J., {Casertano}, S., {et~al.} 2022, arXiv e-prints,
  arXiv:2201.13296

\bibitem[{Silverwood \& Easther(2019)}]{silverwood_easther2019}
Silverwood, H., \& Easther, R. 2019, Publications of the Astronomical Society
  of Australia, 36, e038

\bibitem[{{Sozzetti} {et~al.}(2014){Sozzetti}, {Giacobbe}, {Lattanzi},
  {Micela}, {Morbidelli}, \& {Tinetti}}]{Sozzetti2014}
{Sozzetti}, A., {Giacobbe}, P., {Lattanzi}, M.~G., {et~al.} 2014, \mnras, 437,
  497

\bibitem[{{Steiger} {et~al.}(2021){Steiger}, {Currie}, {Brandt}, {Guyon},
  {Kuzuhara}, {Chilcote}, {Groff}, {Lozi}, {Walter}, {Fruitwala}, {Bailey},
  {Zobrist}, {Swimmer}, {Lipartito}, {Smith}, {Bockstiegel}, {Meeker},
  {Coiffard}, {Dodkins}, {Szypryt}, {Davis}, {Daal}, {Bumble}, {Vievard},
  {Sahoo}, {Deo}, {Jovanovic}, {Martinache}, {Doppmann}, {Tamura}, {Kasdin}, \&
  {Mazin}}]{steiger2021}
{Steiger}, S., {Currie}, T., {Brandt}, T.~D., {et~al.} 2021, \aj, 162, 44

\bibitem[{{Tian} {et~al.}(2019){Tian}, {El-Badry}, {Rix}, \& {Gould}}]{tian19}
{Tian}, H.-J., {El-Badry}, K., {Rix}, H.-W., \& {Gould}, A. 2019, arXiv
  e-prints, arXiv:1909.04765

\bibitem[{{Tisserand} {et~al.}(2007){Tisserand}, {Le Guillou}, {Afonso},
  {Albert}, {Andersen}, {Ansari}, {Aubourg}, {Bareyre}, {Beaulieu}, {Charlot},
  {Coutures}, {Ferlet}, {Fouqu{\'e}}, {Glicenstein}, {Goldman}, {Gould},
  {Graff}, {Gros}, {Haissinski}, {Hamadache}, {de Kat}, {Lasserre}, {Lesquoy},
  {Loup}, {Magneville}, {Marquette}, {Maurice}, {Maury}, {Milsztajn}, {Moniez},
  {Palanque-Delabrouille}, {Perdereau}, {Rahal}, {Rich}, {Spiro},
  {Vidal-Madjar}, {Vigroux}, {Zylberajch}, \& {EROS-2
  Collaboration}}]{tisserand07}
{Tisserand}, P., {Le Guillou}, L., {Afonso}, C., {et~al.} 2007, \aap, 469, 387

\bibitem[{{Tomsick} \& {Muterspaugh}(2010)}]{Tomsick2010}
{Tomsick}, J.~A., \& {Muterspaugh}, M.~W. 2010, \apj, 719, 958

\bibitem[{{Vasiliev}(2019)}]{vasiliev2019}
{Vasiliev}, E. 2019, \mnras, 484, 2832

\bibitem[{{Vigna-G{\'o}mez} \& {Ramirez-Ruiz}(2022)}]{vigna-gomez2022}
{Vigna-G{\'o}mez}, A., \& {Ramirez-Ruiz}, E. 2022, arXiv e-prints,
  arXiv:2203.08478

\bibitem[{Virtanen {et~al.}(2020)Virtanen, Gommers, Oliphant, Haberland, Reddy,
  Cournapeau, Burovski, Peterson, Weckesser, Bright, {van der Walt}, Brett,
  Wilson, Millman, Mayorov, Nelson, Jones, Kern, Larson, Carey, Polat, Feng,
  Moore, {VanderPlas}, Laxalde, Perktold, Cimrman, Henriksen, Quintero, Harris,
  Archibald, Ribeiro, Pedregosa, {van Mulbregt}, \& {SciPy 1.0
  Contributors}}]{scipy}
Virtanen, P., Gommers, R., Oliphant, T.~E., {et~al.} 2020, Nature Methods, 17,
  261

\bibitem[{{Webb} {et~al.}(2012){Webb}, {Cseh}, {Lenc}, {Godet}, {Barret},
  {Corbel}, {Farrell}, {Fender}, {Gehrels}, \& {Heywood}}]{webb2012}
{Webb}, N., {Cseh}, D., {Lenc}, E., {et~al.} 2012, Science, 337, 554

\bibitem[{{Weinberg} {et~al.}(1987){Weinberg}, {Shapiro}, \&
  {Wasserman}}]{weinberg87}
{Weinberg}, M.~D., {Shapiro}, S.~L., \& {Wasserman}, I. 1987, \apj, 312, 367

\bibitem[{{Wielen}(1997)}]{Wielen1997}
{Wielen}, R. 1997, \aap, 325, 367

\bibitem[{{Wyrzykowski} \& {Mandel}(2020)}]{wyrzykowski2020}
{Wyrzykowski}, {\L}., \& {Mandel}, I. 2020, \aap, 636, A20

\bibitem[{{Yalinewich} {et~al.}(2018){Yalinewich}, {Beniamini}, {Hotokezaka},
  \& {Zhu}}]{Yalinewich2018}
{Yalinewich}, A., {Beniamini}, P., {Hotokezaka}, K., \& {Zhu}, W. 2018, \mnras,
  481, 930

\bibitem[{{Yamaguchi} {et~al.}(2018){Yamaguchi}, {Kawanaka}, {Bulik}, \&
  {Piran}}]{Yamaguchi2018}
{Yamaguchi}, M.~S., {Kawanaka}, N., {Bulik}, T., \& {Piran}, T. 2018, \apj,
  861, 21

\bibitem[{{Yoo} {et~al.}(2004){Yoo}, {Chanam{\'e}}, \& {Gould}}]{yoo04}
{Yoo}, J., {Chanam{\'e}}, J., \& {Gould}, A. 2004, \apj, 601, 311

\bibitem[{{Zeng} {et~al.}(2021){Zeng}, {Brandt}, {Li}, {Dupuy}, {Li}, {Brandt},
  {Farihi}, {Horner}, {Wittenmyer}, {Paul. Butler}, {Tinney}, {Carter},
  {Wright}, {Jones}, \& {O'Toole}}]{zeng2021}
{Zeng}, Y., {Brandt}, T.~D., {Li}, G., {et~al.} 2021, arXiv e-prints,
  arXiv:2112.06394

\end{thebibliography}

\end{document}